\documentclass[twocolumn]{revtex4}
\usepackage{amsfonts}
\usepackage{amsmath}
\usepackage{graphicx}
\usepackage{dcolumn}
\usepackage{bm}
\usepackage{amssymb}

\begin{document}

\title{Quantum-dot thermometry}
\author{E.A. Hoffmann}
\author{N. Nakpathomkun}
\author{A.I. Persson}
\author{H. Linke}
\email{linke@uoregon.edu}
\affiliation{Physics Dept. and Materials Science Institute, University of Oregon, Eugene,
OR 97403-1274, USA}
\author{H.A. Nilsson}
\author{L. Samuelson}
\affiliation{Solid State Physics/The Nanometer Structure Consortium, Lund University, Box
118, S-221 00, Lund, Sweden}

\begin{abstract}
We present a method for the measurement of a temperature differential across
a single quantum dot that has transmission resonances that are separated in
energy by much more than the thermal energy. We determine numerically that
the method is accurate to within a few percent across a wide range of
parameters. The proposed method measures the temperature of the electrons
that enter the quantum dot and will be useful in experiments that aim to
test theory which predicts quantum dots are highly efficient thermoelectrics.
\end{abstract}

\maketitle
In the ongoing development of effective thermoelectric materials and
devices, low-dimensional systems are of particularly great interest, because
to optimize the performance of a thermoelectric, it is crucial to control
the energy spectrum of mobile electrons \cite{Hicks03(2), Mahan96,
Humphrey02, Humphrey05, Humphrey05PRL}. Devices for high-efficiency
thermal-to-electric power conversion based on quantum dots defined by double
barriers embedded in nanowires have been proposed \cite{O'Dwyer06}. Such
systems have great advantages, because they select the energies at which
electrons are transmitted \cite{Bjork02, Bjork04}, and because nanowires can
be contacted in highly ordered arrays \cite{Bryllert06} with the potential
for large-scale parallel operation.

In order to measure quantitatively the dependence of thermopower and
energy-conversion efficiency on the transmission spectrum of a quantum dot,
it is necessary to apply and determine accurately a temperature differential
across the dot. Traditionally for the thermoelectric characterization of
mesoscopic devices such as quantum point contacts \cite{Molenkamp92},
quantum dots in 2DEG's \cite{Molenkamp94}, carbon nanotubes \cite{Hone98,
Kim03, Llaguno04}, and nanowires \cite{Seol07, Shi03}, an ac heating current
generates a temperature differential that is measured in separate
calibration experiments. Here we propose a technique that measures the
actual \textit{electronic} temperature differential across a quantum dot and
does not require separate calibration. The basic concept is as follows: the
change in current across a quantum dot in response to an applied heating
voltage, $V_{\text{H}}$, is measured. This signal contains information about
the electron temperatures at the source and drain, but it also depends on
the dot's energy-dependent transmission function, $\tau \left( \varepsilon
\right) $. However, one can obtain the necessary information about $\tau
\left( \varepsilon \right) $ from conductance measurements. Together, these
two measurements allow one to determine the source and drain temperatures
separately.

The two-terminal current through a quantum dot can be written \cite
{Landauer57, Landauer70}
\begin{equation}
I=\frac{2e}{h}\int_{-\infty }^{\infty }\left[ f_{\text{s}}\left( \varepsilon
\right) -f_{\text{d}}\left( \varepsilon \right) \right] \tau \left(
\varepsilon \right) d\varepsilon ,  \label{Landauer}
\end{equation}
where $f_{\text{s,d}}^{-1}=e^{\xi _{\text{s,d}}}+1$ are the Fermi-Dirac
distributions in the nanowire's source and drain leads, respectively, and
their arguments are $\xi _{\text{s,d}}=\left( \varepsilon -\mu _{\text{s,d}
}\mp eV/2\right) /k_{\text{B}}T_{\text{s,d}}$. We assume the bias voltage, $
V $, is applied symmetrically across the dot. For the case of a quantum dot
or single-electron transistor (SET) with well-separated transmission maxima
as a function of gate voltage, Eq.~(\ref{Landauer}) predicts the
characteristic Coulomb blockade diamonds which appear in the differential
conductance, $G$, as a function of bias voltage and gate voltage. 
\begin{figure}[ht]
\begin{center}
\leavevmode
\includegraphics{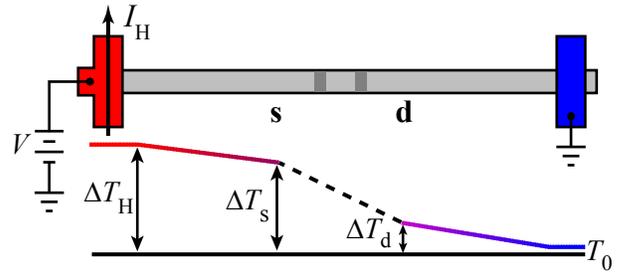}
\end{center}
\caption{The heating setup and the temperature landscape. The source contact
is warmed with a voltage-balanced heating current and can be biased for
thermocurrent measurements. Electron transport through the quantum dot is
determined by the local temperatures of the source and drain sides of the
dot, $T_{\text{ s,d}}=\Delta T_{\text{s,d}}+T_{ \text{0}}$. (Color online)}
\label{setup}
\end{figure}

In a typical experiment, an ac heating current is used to modulate the
temperature $T_{\text{H}}$ of an ohmic contact at one end of a nanowire
(taken here to be the source contact, Fig.~\ref{setup})\cite{footnote1} 
with amplitude $\Delta T_{\text{H}}$ with respect to the
unperturbed device temperature, $T_{\text{0}}$. We are interested in the
associated electronic temperature rises, $\Delta T_{\text{s,d}}=T_{\text{s,d}
}-T_{0}$, in the immediate vicinity of the quantum dot 
(see Fig.~\ref{setup}). In the case of strong electron- phonon interaction (for example near room
temperature) the electronic temperature will drop linearly along the
nanowire, and $\Delta T_{\text{s}}\cong \Delta T_{\text{d}}$ if the quantum
dot is short compared to the wire. At low temperatures, however, where
electron-phonon interaction in the nanowire is expected to be weak, $\Delta
T_{\text{H}}>\Delta T_{\text{s }}>\Delta T_{\text{d}}>0$, and $\Delta T_{
\text{s}}$ and $\Delta T_{\text{ d}}$ need to be measured.

Assuming an ac heating voltage, $V_{\text{H}}=V_{\text{0}}\cos (\omega t$),
the temperature rises on the source and drain sides of the dot can be
written $\Delta T_{\text{s,d}}=\beta _{\text{s,d}}V_{\text{H}}^{\gamma }$ ,
where $\beta _{\text{s,d}}$ are unknown constants and $\gamma $ can take on
various values depending on the type and strength of electron-phonon
interaction in the heating wire \cite{Henny97}. Here we assume a short
heating wire and Joule heating, and therefore $\gamma =2$ \cite{Henny97}. In
this regime, by an application of the chain rule, the \textit{rms}-amplitude
of the ac temperature rises can be written 
\begin{equation}
\Delta T_{\text{s,d}}\cong V_{0}^{2}\frac{\partial T_{\text{s,d }}}{\partial
\left( V_{\text{H}}^{2}\right) }=V_{0}^{2}\left( \frac{\partial I}{\partial
T_{\text{s,d}}}\right) ^{-1}\frac{\partial I}{\partial \left( V_{\text{H}
}^{2}\right) }.  \label{key relationship}
\end{equation}
In an experiment, one can measure $\partial I/\partial \left( V_{\text{H}
}^{2}\right) $, the frequency-doubled response to the ac heating voltage.
The differential thermocurrent, 
\begin{equation}
\frac{\partial I}{\partial T_{\text{s,d}}}=\frac{2e}{h}\int_{-\infty
}^{\infty }\left( \mp \frac{\partial f_{\text{s,d}}}{\partial \xi _{\text{
s,d }}}\frac{\xi _{\text{s,d}}}{T_{\text{s,d}}}\right) \tau \left(
\varepsilon \right) d\varepsilon ,  \label{Temp final}
\end{equation}
cannot be measured directly. \ However, we will show that it can be obtained
in good approximation from conductance measurements.

Under bias conditions, where the source (drain) electrochemical potential is
near a well-defined transmission resonance of the quantum dot, while the
drain (source) is several $k_{\text{B}}T$ away from the next resonance, the
second derivative of the current is 
\begin{equation}
\frac{\partial ^{2}I}{\partial V^{2}}\cong \frac{e^{2}}{4k_{\text{B}}^{2}} 
\frac{1}{T_{\text{s,d}}}\frac{2e}{h}\int_{-\infty }^{\infty }\left( \pm 
\frac{\partial f_{\text{s,d}}}{\partial \xi _{\text{s,d}}}\frac{2f_{\text{
s,d }}-1}{T_{\text{s,d}}}\right) \tau \left( \varepsilon \right)
d\varepsilon .  \label{d2idv2}
\end{equation}
A key observation is that the integrands in Eq.~(\ref{Temp final}) and Eq.~(
\ref{d2idv2}) are qualitatively very similar: 
\begin{equation}
\frac{\partial f_{\text{s,d}}}{\partial \xi _{\text{s,d}}}\frac{2f_{\text{
s,d }}-1}{T_{\text{s,d}}}\cong -\frac{1}{2}\frac{\partial f_{\text{s,d}}}{
\partial \xi _{\text{s,d}}}\frac{\xi _{\text{s,d}}}{T_{\text{s,d}}}.
\label{approx}
\end{equation}
This approximation holds for all $\xi _{\text{s,d}}$, because $2f_{\text{s,d}
}-1$ limits to $-\xi _{\text{s,d}}/2$ when $\xi _{\text{s,d}}$ is small, and 
$\partial f_{\text{s,d}}/\partial \xi _{\text{s,d}}$ goes to zero in all
other cases. \ With this approximation, we can combine Eqs.~(\ref{Temp final}
) and (\ref{d2idv2}): 
\begin{equation}
\frac{\partial I}{\partial T_{\text{s,d}}}\cong \left( \Lambda \frac{e^{2}}{
4k_{\text{B}}^{2}}\frac{1}{T_{\text{s,d}}}\right) ^{-1}\frac{\partial ^{2}I}{
\partial V^{2}},  \label{approx2}
\end{equation}
where $\Lambda $ is a unitless scaling factor introduced during integration.
In this way, all the information about $\tau \left( \varepsilon \right) $
needed to determine $\partial I/\partial T_{\text{s,d}}$ is accounted for by
measuring $\partial ^{2}I/\partial V^{2}$. 
\begin{figure}[t]
\begin{center}
\leavevmode
\includegraphics{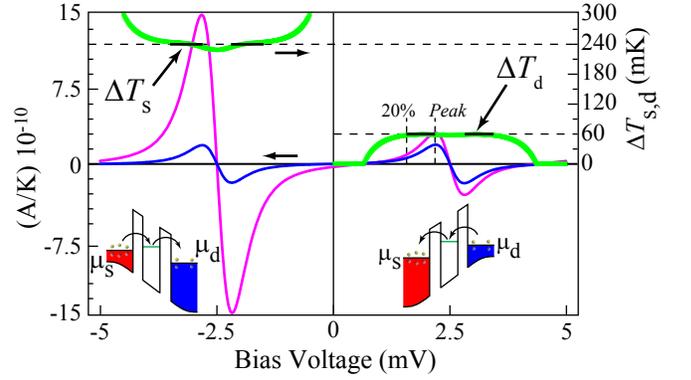}
\end{center}
\caption{A plot of $\partial I/\partial T_{H}$ (red) and $(2k_{\text{B}}/e)^{2}\Delta T_{\text{H}}\partial
^{2}I/\partial V^{2}$ (blue) as a function of bias voltage at a fixed gate
voltage, as indicated by the horizontal dashed line in Fig.~\ref{3D}. 
The temperature rise (green) is calculated via
Eq.~(\ref{final}). Insets: The position of the resonant tunneling
energy of the dot with respect to the electrochemical potentials in the
leads. In this model, we used a transmission function consisting of
Lorentzians with a FWHM of $\Gamma =0.5$ meV equally spaced by 5 meV and $
\Delta T_{\text{H} }=300$ mK, $\Delta T_{\text{s}}=0.8\Delta T_{\text{H}}$ $
=240$ mK, $\Delta T_{\text{d}}=0.2\Delta T_{\text{H}}$ $=60$ mK, and $T_{
\text{0}}=230$ mK. (Color online)}
\label{Slice}
\end{figure}
\begin{figure}[ht]
\begin{center}
\leavevmode\includegraphics{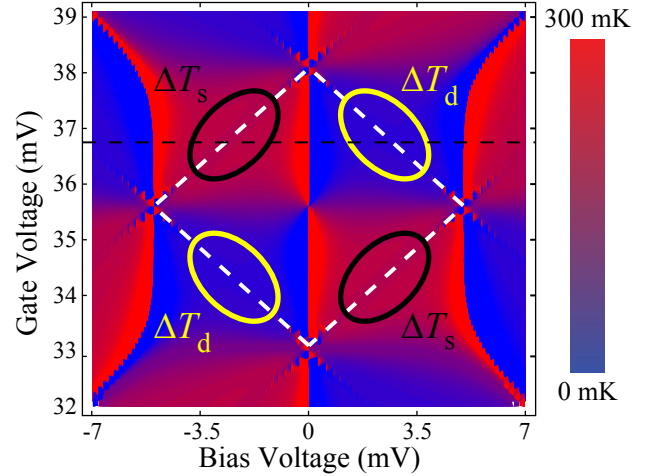}
\end{center}
\caption{The calculated temperature rise as a function of bias and gate
voltages for a single Coulomb blockade diamond. In the regions indicated,
either the source or drain electrochemical potential, but not both, is
within a few $k_{\text{B}}T$ of a resonant energy of the dot. Our
assumptions are fulfilled in these regions, and the simulation produces
temperature plateaus predicting the correct $\Delta T_{\text{s,d}}$. 
Fig.~\ref{Slice} shows a slice at 36.75 mV gate voltage, as indicated by
the dashed line. (Color online)}
\label{3D}
\end{figure}

Substituting Eq.~(\ref{approx2}) into Eq.~(\ref{key relationship}) and
solving for $\Delta T_{\text{s,d}}$ yields our final result, 
\begin{equation}
\Delta T_{\text{s,d}}=\dfrac{1}{2}\sqrt{T_{\text{0}}^{2}+\Lambda \dfrac{
e^{2} }{k_{\text{B}}^{2}}V_{\text{0}}^{2}\left( \frac{\partial ^{2}I}{
\partial V^{2}}\right) ^{-1}\dfrac{\partial I}{\partial \left( V_{\text{H}
}^{2}\right) }}-\dfrac{T_{\text{0}}}{2},  \label{final}
\end{equation}
which shows that approximations of $\Delta T_{\text{s}}$ and $\Delta T_{ 
\text{d}}$ can be obtained from measurement of $\partial ^{2}I/\partial
V^{2} $ and $\partial I/\partial \left( V_{\text{H}}^{2}\right) $ and
knowledge of $T_{0}$.

To illustrate the qualitative similarity of Eqs.~(\ref{Temp final}) and (\ref
{d2idv2}), we show numerical calculations of the two in Fig.~\ref{Slice}
taken at the gate voltage indicated by the horizontal dashed line in 
Fig.~\ref{3D} at 36.75 mV. The left inset of Fig.~\ref{Slice} illustrates that, in this example, when the bias voltage is negative, the source temperature
is the only temperature affecting the current through the dot; therefore, $
\partial I/ \partial T_{\text{H}}=\partial I/\partial T_{\text{s}}$ in this
bias configuration. In the opposite configuration, $\partial I/\partial T_{
\text {H}}=\partial I/\partial T_{\text{d}}$, as shown in the right inset of
Fig.~\ref{Slice}. Fig.~\ref{3D} shows Eq.~(\ref{final}) as calculated from modeled data of
Eqs.~(\ref{Temp final}) and (\ref{d2idv2}) across an entire Coulomb blockade
diamond (indicated by a white, dashed line), and a slice through that
diamond is shown as green symbols in Fig.~\ref{Slice}. In regions along the
diamond ridges (circled areas in Fig.~\ref{3D})---where one, and only one,
of the two electrochemical potentials in the source or drain is within a few 
$k_{\text{B}}T$ of a transmission resonance---Eq.~(\ref{final}) yields
consistent values in accordance with the assumptions that allow us to write
Eq.~(\ref{d2idv2}). In all other regions, Eq.~(\ref{d2idv2}) is not valid,
not even approximately, because it only accounts for one Fermi-Dirac
distribution.
\begin{figure}[t]
\begin{center}
\leavevmode\includegraphics{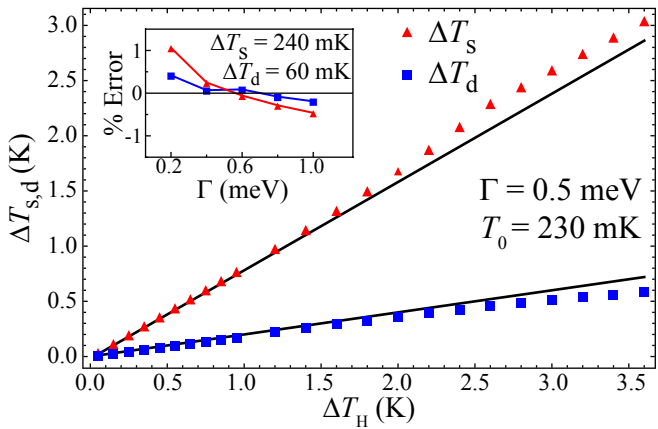}
\end{center}
\caption{The calculated temperature rises as a function of $\Delta T_{\text{
H }}.$ The calculated values agree well with the expected values (solid
lines) up to nearly 10~$T_{\text{0}}.$ Inset: The percent error as a
function of $\Gamma $, the full width at half max (FWHM) of the transmission
function, $\tau \left(\varepsilon \right) $ in Eq.~(\ref{Landauer}). The error is within 1\% over an order of magnitude in $\Gamma $. Here $\Delta T_{\text{s}}$ is 240 mK and $\Delta T_{ \text{d}}$ is
60 mK. (Color online)}
\label{error}
\end{figure}

The use of this method requires knowledge of the appropriate scaling factor 
$\Lambda $, defined in Eq.~\ref{approx2}, which needs to be determined numerically. For the particular
modeling parameters used here (see caption of Fig.~\ref{Slice}), we found 
$\Lambda=0.304$ by averaging Eq.~(\ref{final}) over the voltage range from
the peak value of $\partial I/\partial T_{\text{H}}$ to 20\% of its peak
value, where the signal-to-noise ratio in an experiment should be largest.
Note that for different parameters, $\Lambda$ will differ, but it is
insensitive to typical experimental variations in $\Gamma$ and $\Delta T_{
\text{H}}$. For example, in Fig.~\ref{error}, we
show that the use of the same $\Lambda =0.304$ (calculated for $\Gamma=0.5$
meV) yields errors in $\Delta T_{\text{s}}$ and $\Delta T_{\text{d}}$ of
only 1\% when $\Gamma $ is varied over nearly an order of magnitude around $
\Gamma=0.5$ meV (inset of Fig.~\ref{error}) and only a few percent for $
\Delta T_{\text {H}}$ up to almost 10 $T_{\text{0}}$. To put this small
error into context, note that the local temperatures $T_{\text{s} } $ and $
T_{\text{d }}$ can be defined only over a distance of about an inelastic
scattering length, such that an accuracy of less than a few percent is not
necessarily physically meaningful.

This research was supported by ONR, ONR Global, the Swedish Research Council
(VR), the Foundation for Strategic Research (SSF), the Knut and Alice
Wallenberg Foundation, and an NSF-IGERT Fellowship.


\begin{thebibliography}{20}
\expandafter\ifx\csname natexlab\endcsname\relax\def\natexlab#1{#1}\fi
\expandafter\ifx\csname bibnamefont\endcsname\relax
  \def\bibnamefont#1{#1}\fi
\expandafter\ifx\csname bibfnamefont\endcsname\relax
  \def\bibfnamefont#1{#1}\fi
\expandafter\ifx\csname citenamefont\endcsname\relax
  \def\citenamefont#1{#1}\fi
\expandafter\ifx\csname url\endcsname\relax
  \def\url#1{\texttt{#1}}\fi
\expandafter\ifx\csname urlprefix\endcsname\relax\def\urlprefix{URL }\fi
\providecommand{\bibinfo}[2]{#2}
\providecommand{\eprint}[2][]{\url{#2}}

\bibitem[{\citenamefont{Hicks and Dresselhaus}(1993)}]{Hicks03(2)}
\bibinfo{author}{\bibfnamefont{L.~D.} \bibnamefont{Hicks}} \bibnamefont{and}
  \bibinfo{author}{\bibfnamefont{M.~S.} \bibnamefont{Dresselhaus}},
  \bibinfo{journal}{Phys. Rev. B} \textbf{\bibinfo{volume}{47}},
  \bibinfo{pages}{12727} (\bibinfo{year}{1993}).

\bibitem[{\citenamefont{Mahan and Sofo}(1996)}]{Mahan96}
\bibinfo{author}{\bibfnamefont{G.~D.} \bibnamefont{Mahan}} \bibnamefont{and}
  \bibinfo{author}{\bibfnamefont{J.~O.} \bibnamefont{Sofo}},
  \bibinfo{journal}{Proc. Natl. Acad. Sci. USA} \textbf{\bibinfo{volume}{93}},
  \bibinfo{pages}{7436} (\bibinfo{year}{1996}).

\bibitem[{\citenamefont{Humphrey et~al.}(2002)\citenamefont{Humphrey, Newbury,
  Taylor, and Linke}}]{Humphrey02}
\bibinfo{author}{\bibfnamefont{T.~E.} \bibnamefont{Humphrey}},
  \bibinfo{author}{\bibfnamefont{R.}~\bibnamefont{Newbury}},
  \bibinfo{author}{\bibfnamefont{R.~P.} \bibnamefont{Taylor}},
  \bibnamefont{and} \bibinfo{author}{\bibfnamefont{H.}~\bibnamefont{Linke}},
  \bibinfo{journal}{Phys. Rev. Lett.} \textbf{\bibinfo{volume}{89}},
  \bibinfo{pages}{116801} (\bibinfo{year}{2002}).

\bibitem[{\citenamefont{Humphrey et~al.}(2005)\citenamefont{Humphrey, O'Dwyer,
  and Linke}}]{Humphrey05}
\bibinfo{author}{\bibfnamefont{T.~E.} \bibnamefont{Humphrey}},
  \bibinfo{author}{\bibfnamefont{M.~F.} \bibnamefont{O'Dwyer}},
  \bibnamefont{and} \bibinfo{author}{\bibfnamefont{H.}~\bibnamefont{Linke}},
  \bibinfo{journal}{J. Phys. D} \textbf{\bibinfo{volume}{38}},
  \bibinfo{pages}{2051} (\bibinfo{year}{2005}).

\bibitem[{\citenamefont{Humphrey and Linke}(2005)}]{Humphrey05PRL}
\bibinfo{author}{\bibfnamefont{T.~E.} \bibnamefont{Humphrey}} \bibnamefont{and}
  \bibinfo{author}{\bibfnamefont{H.}~\bibnamefont{Linke}},
  \bibinfo{journal}{Phys. Rev. Lett.} \textbf{\bibinfo{volume}{94}},
  \bibinfo{pages}{096601} (\bibinfo{year}{2005}).

\bibitem[{\citenamefont{O'Dwyer et~al.}(2006)\citenamefont{O'Dwyer, Humphrey,
  and Linke}}]{O'Dwyer06}
\bibinfo{author}{\bibfnamefont{M.~F.} \bibnamefont{O'Dwyer}},
  \bibinfo{author}{\bibfnamefont{T.~E.} \bibnamefont{Humphrey}},
  \bibnamefont{and} \bibinfo{author}{\bibfnamefont{H.}~\bibnamefont{Linke}},
  \bibinfo{journal}{Nanotech.} \textbf{\bibinfo{volume}{17}},
  \bibinfo{pages}{S338} (\bibinfo{year}{2006}).

\bibitem[{\citenamefont{Bj\"{o}rk et~al.}(2002)\citenamefont{Bj\"{o}rk,
  Ohlsson, Sass, Persson, Thelander, Magnusson, Deppert, Wallenberg, and
  Samuelson}}]{Bjork02}
\bibinfo{author}{\bibfnamefont{M.~T.} \bibnamefont{Bj\"{o}rk}},
  \bibinfo{author}{\bibfnamefont{B.~J.} \bibnamefont{Ohlsson}},
  \bibinfo{author}{\bibfnamefont{T.}~\bibnamefont{Sass}},
  \bibinfo{author}{\bibfnamefont{A.~I.} \bibnamefont{Persson}},
  \bibinfo{author}{\bibfnamefont{C.}~\bibnamefont{Thelander}},
  \bibinfo{author}{\bibfnamefont{M.~H.} \bibnamefont{Magnusson}},
  \bibinfo{author}{\bibfnamefont{K.}~\bibnamefont{Deppert}},
  \bibinfo{author}{\bibfnamefont{L.~R.} \bibnamefont{Wallenberg}},
  \bibnamefont{and}
  \bibinfo{author}{\bibfnamefont{L.}~\bibnamefont{Samuelson}},
  \bibinfo{journal}{Nano Lett.} \textbf{\bibinfo{volume}{2}},
  \bibinfo{pages}{87} (\bibinfo{year}{2002}).

\bibitem[{\citenamefont{Bj\"{o}rk et~al.}(2004)\citenamefont{Bj\"{o}rk,
  Thelander, Hansen, Jensen, Larsson, Wallenberg, and Samuelson}}]{Bjork04}
\bibinfo{author}{\bibfnamefont{M.~T.} \bibnamefont{Bj\"{o}rk}},
  \bibinfo{author}{\bibfnamefont{C.}~\bibnamefont{Thelander}},
  \bibinfo{author}{\bibfnamefont{A.~E.} \bibnamefont{Hansen}},
  \bibinfo{author}{\bibfnamefont{L.~E.} \bibnamefont{Jensen}},
  \bibinfo{author}{\bibfnamefont{M.~W.} \bibnamefont{Larsson}},
  \bibinfo{author}{\bibfnamefont{L.~R.} \bibnamefont{Wallenberg}},
  \bibnamefont{and}
  \bibinfo{author}{\bibfnamefont{L.}~\bibnamefont{Samuelson}},
  \bibinfo{journal}{Nano Lett.} \textbf{\bibinfo{volume}{4}},
  \bibinfo{pages}{1621} (\bibinfo{year}{2004}).

\bibitem[{\citenamefont{Bryllert et~al.}(2006)\citenamefont{Bryllert,
  Wernersson, L\"{o}wgren, and Samuelson}}]{Bryllert06}
\bibinfo{author}{\bibfnamefont{T.}~\bibnamefont{Bryllert}},
  \bibinfo{author}{\bibfnamefont{L.-E.} \bibnamefont{Wernersson}},
  \bibinfo{author}{\bibfnamefont{T.}~\bibnamefont{L\"{o}wgren}},
  \bibnamefont{and}
  \bibinfo{author}{\bibfnamefont{L.}~\bibnamefont{Samuelson}},
  \bibinfo{journal}{Nanotech.} \textbf{\bibinfo{volume}{17}},
  \bibinfo{pages}{S227} (\bibinfo{year}{2006}).

\bibitem[{\citenamefont{Molenkamp et~al.}(1992)\citenamefont{Molenkamp,
  Gravier, van Houten, Buijk, Mabesoone, and Foxon}}]{Molenkamp92}
\bibinfo{author}{\bibfnamefont{L.~W.} \bibnamefont{Molenkamp}},
  \bibinfo{author}{\bibfnamefont{T.}~\bibnamefont{Gravier}},
  \bibinfo{author}{\bibfnamefont{H.}~\bibnamefont{van Houten}},
  \bibinfo{author}{\bibfnamefont{O.~J.~A.} \bibnamefont{Buijk}},
  \bibinfo{author}{\bibfnamefont{M.~A.~A.} \bibnamefont{Mabesoone}},
  \bibnamefont{and} \bibinfo{author}{\bibfnamefont{C.~T.} \bibnamefont{Foxon}},
  \bibinfo{journal}{Phys. Rev. Lett.} \textbf{\bibinfo{volume}{68}},
  \bibinfo{pages}{3765} (\bibinfo{year}{1992}).

\bibitem[{\citenamefont{Molenkamp et~al.}(1994)\citenamefont{Molenkamp,
  Staring, Alphenaar, van Houten, and Beenakker}}]{Molenkamp94}
\bibinfo{author}{\bibfnamefont{L.~W.} \bibnamefont{Molenkamp}},
  \bibinfo{author}{\bibfnamefont{A.~A.~M.} \bibnamefont{Staring}},
  \bibinfo{author}{\bibfnamefont{B.~W.} \bibnamefont{Alphenaar}},
  \bibinfo{author}{\bibfnamefont{H.}~\bibnamefont{van Houten}},
  \bibnamefont{and} \bibinfo{author}{\bibfnamefont{C.~W.~J.}
  \bibnamefont{Beenakker}}, \bibinfo{journal}{Semicond. Sci. Technol.}
  \textbf{\bibinfo{volume}{9}}, \bibinfo{pages}{903} (\bibinfo{year}{1994}).

\bibitem[{\citenamefont{Hone et~al.}(1998)\citenamefont{Hone, Ellwood, Muno,
  Mizel, Cohen, Zettl, Rinzler, and Smalley}}]{Hone98}
\bibinfo{author}{\bibfnamefont{J.}~\bibnamefont{Hone}},
  \bibinfo{author}{\bibfnamefont{I.}~\bibnamefont{Ellwood}},
  \bibinfo{author}{\bibfnamefont{M.}~\bibnamefont{Muno}},
  \bibinfo{author}{\bibfnamefont{A.}~\bibnamefont{Mizel}},
  \bibinfo{author}{\bibfnamefont{M.~L.} \bibnamefont{Cohen}},
  \bibinfo{author}{\bibfnamefont{A.}~\bibnamefont{Zettl}},
  \bibinfo{author}{\bibfnamefont{A.~G.} \bibnamefont{Rinzler}},
  \bibnamefont{and} \bibinfo{author}{\bibfnamefont{R.~E.}
  \bibnamefont{Smalley}}, \bibinfo{journal}{Phys. Rev. Lett.}
  \textbf{\bibinfo{volume}{80}}, \bibinfo{pages}{1042} (\bibinfo{year}{1998}).

\bibitem[{\citenamefont{Small et~al.}(2003)\citenamefont{Small, Perez, and
  Kim}}]{Kim03}
\bibinfo{author}{\bibfnamefont{J.~P.} \bibnamefont{Small}},
  \bibinfo{author}{\bibfnamefont{K.~M.} \bibnamefont{Perez}}, \bibnamefont{and}
  \bibinfo{author}{\bibfnamefont{P.}~\bibnamefont{Kim}},
  \bibinfo{journal}{Phys. Rev. Lett.} \textbf{\bibinfo{volume}{91}},
  \bibinfo{pages}{256801} (\bibinfo{year}{2003}).

\bibitem[{\citenamefont{Llaguno et~al.}(2004)\citenamefont{Llaguno, Fischer,
  Johnson, and Hone}}]{Llaguno04}
\bibinfo{author}{\bibfnamefont{M.~C.} \bibnamefont{Llaguno}},
  \bibinfo{author}{\bibfnamefont{J.~E.} \bibnamefont{Fischer}},
  \bibinfo{author}{\bibfnamefont{A.~T.} \bibnamefont{Johnson}},
  \bibnamefont{and} \bibinfo{author}{\bibfnamefont{J.}~\bibnamefont{Hone}},
  \bibinfo{journal}{Nano Lett.} \textbf{\bibinfo{volume}{4}},
  \bibinfo{pages}{45} (\bibinfo{year}{2004}).

\bibitem[{\citenamefont{Shi et~al.}(2003)\citenamefont{Shi, Li, Yu, Jang, Kim,
  Yao, Kim, and Majumdar}}]{Shi03}
\bibinfo{author}{\bibfnamefont{L.}~\bibnamefont{Shi}},
  \bibinfo{author}{\bibfnamefont{D.~Y.} \bibnamefont{Li}},
  \bibinfo{author}{\bibfnamefont{C.~H.} \bibnamefont{Yu}},
  \bibinfo{author}{\bibfnamefont{W.~Y.} \bibnamefont{Jang}},
  \bibinfo{author}{\bibfnamefont{D.}~\bibnamefont{Kim}},
  \bibinfo{author}{\bibfnamefont{Z.}~\bibnamefont{Yao}},
  \bibinfo{author}{\bibfnamefont{P.}~\bibnamefont{Kim}}, \bibnamefont{and}
  \bibinfo{author}{\bibfnamefont{A.}~\bibnamefont{Majumdar}},
  \bibinfo{journal}{J. Heat Trans.} \textbf{\bibinfo{volume}{125}},
  \bibinfo{pages}{881} (\bibinfo{year}{2003}).

\bibitem[{\citenamefont{Seol et~al.}(2007)\citenamefont{Seol, Moore, Saha,
  Zhou, Shi, Ye, Scheffler, Mingo, and Yamada}}]{Seol07}
\bibinfo{author}{\bibfnamefont{J.~H.} \bibnamefont{Seol}},
  \bibinfo{author}{\bibfnamefont{A.~L.} \bibnamefont{Moore}},
  \bibinfo{author}{\bibfnamefont{S.~K.} \bibnamefont{Saha}},
  \bibinfo{author}{\bibfnamefont{F.}~\bibnamefont{Zhou}},
  \bibinfo{author}{\bibfnamefont{L.}~\bibnamefont{Shi}},
  \bibinfo{author}{\bibfnamefont{Q.~L.} \bibnamefont{Ye}},
  \bibinfo{author}{\bibfnamefont{R.}~\bibnamefont{Scheffler}},
  \bibinfo{author}{\bibfnamefont{N.}~\bibnamefont{Mingo}}, \bibnamefont{and}
  \bibinfo{author}{\bibfnamefont{T.}~\bibnamefont{Yamada}},
  \bibinfo{journal}{J. Appl. Phys.} \textbf{\bibinfo{volume}{101}},
  \bibinfo{pages}{023706} (\bibinfo{year}{2007}).

\bibitem[{\citenamefont{Landauer}(1957)}]{Landauer57}
\bibinfo{author}{\bibfnamefont{R.}~\bibnamefont{Landauer}},
  \bibinfo{journal}{IBM J. Res. Dev} \textbf{\bibinfo{volume}{1}},
  \bibinfo{pages}{223} (\bibinfo{year}{1957}).

\bibitem[{\citenamefont{Landauer}(1970)}]{Landauer70}
\bibinfo{author}{\bibfnamefont{R.}~\bibnamefont{Landauer}},
  \bibinfo{journal}{Phil. Mag.} \textbf{\bibinfo{volume}{21}},
  \bibinfo{pages}{863} (\bibinfo{year}{1970}).

\bibitem[{foo()}]{footnote1}
\bibinfo{note}{Two ac heating voltages, out of phase from one another by
  $180^{\circ }$, are applied to the vertical leads of the source ohmic contact
  (see Fig.~\ref{setup}). These voltages are then tuned so that their sum is
  zero where the contact and nanowire intersect. In this way, the heating
  voltage does not interfere with thermoelectric measurements.}

\bibitem[{\citenamefont{Henny et~al.}(1997)\citenamefont{Henny, Birk, Huber,
  Strunk, Bachtold, Kr\"{u}ger, and Sch\"{o}nenberger}}]{Henny97}
\bibinfo{author}{\bibfnamefont{M.}~\bibnamefont{Henny}},
  \bibinfo{author}{\bibfnamefont{H.}~\bibnamefont{Birk}},
  \bibinfo{author}{\bibfnamefont{R.}~\bibnamefont{Huber}},
  \bibinfo{author}{\bibfnamefont{C.}~\bibnamefont{Strunk}},
  \bibinfo{author}{\bibfnamefont{A.}~\bibnamefont{Bachtold}},
  \bibinfo{author}{\bibfnamefont{M.}~\bibnamefont{Kr\"{u}ger}},
  \bibnamefont{and}
  \bibinfo{author}{\bibfnamefont{C.}~\bibnamefont{Sch\"{o}nenberger}},
  \bibinfo{journal}{Appl. Phys. Lett.} \textbf{\bibinfo{volume}{71}},
  \bibinfo{pages}{773} (\bibinfo{year}{1997}).

\end{thebibliography}
\end{document}